\documentclass[prb,twocolumn,notitlepage,longbibliography]{revtex4}
\usepackage[cp1251]{inputenc}
\usepackage{amsmath}
\usepackage{amssymb}
\usepackage{hyperref}
\usepackage{color}
\usepackage{natbib}
\hypersetup{pdfstartview={FitH},pdfpagemode={UseNone},
            colorlinks,linkcolor=blue, citecolor=blue, urlcolor=blue,
            bookmarks=true, bookmarksopen=true, pdfnewwindow=true}

\usepackage{bm}
\usepackage{color}
\usepackage{epsfig}

\begin{document}
 \renewcommand {\Im}{\mathop\mathrm{Im}\nolimits}
  \renewcommand {\Re}{\mathop\mathrm{Re}\nolimits}
\renewcommand {\i}{{\rm i}}
\newcommand {\sign}{\mathop{\mathrm{sign}}\nolimits}
\newcommand {\e}{{\rm e}}
\newcommand {\rmd}{{\rm d}}
\renewcommand {\phi}{\varphi}
\renewcommand {\epsilon}{\varepsilon}
\newcommand {\eps}{\varepsilon}
\newcommand{\nix}[1]{}
\title{Nonradiative and radiative F\"orster energy transfer between quantum dots}
 \author{Alexander N. Poddubny and Anna V. Rodina}
 \affiliation{Ioffe  Institute, St Petersburg 194021, Russia}
 \email{poddubny@coherent.ioffe.ru}
\begin{abstract}We study theoretically  nonradiative and  radiative  energy transfer between two localized quantum emitters, donor one (i.e. initially excited) and acceptor one (i.e. receiving the excitation). The rates of nonradiative and radiative processes are calculated depending on the spatial and spectral separation between donor and acceptor states and for different donor and  acceptor  lifetimes for typical parameters of  semiconductor quantum dots. We find that the donor lifetime can be significantly modified only due to the nonradiative  F\"orster energy transfer process at donor-acceptor separations $\sim 10$~nm (depending on the acceptor radiative lifetime) and for the energy detuning not larger than 1$\div$2 meV. The efficiency of the  nonradiative  F\"orster energy transfer process under these conditions is close to unity and decreases rapidly with the increase of donor-acceptor distance or energy detuning. At large donor-acceptor separations $>40~$nm the radiative  corrections to the donor lifetime are comparable with nonradiative ones but are relatively weak.
\end{abstract}
\date{\today}
\maketitle
%%%%%%%%%%%%%%%%%%%%%%%%%%%%%%%%%%%%%
\section{Introduction}

\begin{figure}[t!]
\centering\includegraphics[width=0.4\textwidth]{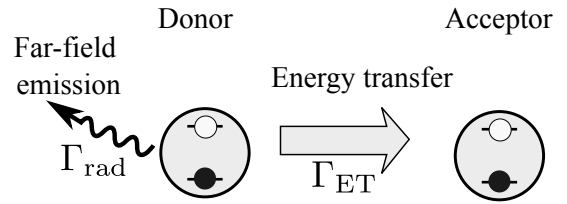}

\caption{Schematic illustration of the energy transfer and photon radiation processes}\label{fig:1}
\end{figure}

F\"orster energy transfer  (ET) processes are now actively studied in various fields that bridge  physics, biology and chemistry. The  energy is transferred from the initially excited (donor) system to the  system that is initially unexcited (acceptor) via the electromagnetic interaction~\cite{Forster1948}. 
This is an incoherent one-way transfer  followed by the rapid emission or nonradiative recombination from the  acceptor state that is to be distinguished from  coherent light-induced coupling.\cite{kulakovskii2006} 
From now on we will use the terms ``donor'' and ``acceptor'' for the energy transmitting and receiving systems. 
Although these terms are quite established in the literature on F\"orster process, they are somewhat ambiguous and should not be confused with donor and acceptor impurities in semiconductor. Here, they characterize excitation transfer and not charge transfer.
 The donor and acceptor systems may be realized as  quantum dots,\cite{Crooker2002,Limpens2015,Rodina2015,Andreakou} quantum wires,\cite{Govorov2013} quantum wells\cite{Agranovich2011,Rindermann} and colloidal nanoplatelets,\cite{Guzelturk2015}, biological molecules,\cite{Fruhwirth,Lopez2014} defects in semiconductor.\cite{DaldossoPRB2009,ProkofievTransferPRB2008}  Typically, the range of the F\"orster interaction is on the order of several nm.\cite{AgranovichGalanin}
By placing the the donors and acceptors into the structured electromagnetic environment one can try to enhance the efficiency of the transfer. In particular, the transfer, mediated by localized and surface plasmons, \cite{Shahbazyan2011,blum2012} photons trapped in the cavity\cite{Andrew2000}  or localized in random glass,\cite{Lopez2014}  as well as modified by metamaterials\cite{Tumkur2015,Zubin2015}  is now actively studied. The concept of tailored photon-induced energy transfer shares a lot of similarities with the Purcell  enhancement\cite{Purcell} of the spontaneous emission in the cavity as compared to that it vacuum. Indeed, in the first case one can think of nonradiative energy transfer from donor to  the acceptor via (virtual) photons, while in the second case the energy is radiated into the real photonic modes (see Fig.~\ref{fig:1}). A   general theory of F\"orster transfer process has been developed in detail.\cite{Andrews1992,Andrews1994,Dung2002,Andrews2004,Klimov2004,delerue_forster_2007}
However, the relation between transfer process and the Purcell effect as well as the character of the transfer in each particular nanosystem, radiative or nonradiative, remains a subject of active discussions.\cite{blum2012,rabouw2014,Tumkur2015,Wubs2015} Simultaneous enhancement and control of the energy transfer and spontaneous emission processes in the same electromagnetic environment are quite challenging.

Here, we study the simplest case of  localized donor and acceptor (e.g. quantum dots), embedded in the dielectric matrix. We first revisit different approaches to calculate the rate of the transfer process and obtain it  consistently  with the donor spontaneous decay rate (Sec.~\ref{sec:model}). Next, we discuss the transfer kinetics (Sec.~\ref{sec:kinetic}) and analyze the radiative and nonradiative contributions to the F\"orster process depending on the spatial and spectral separation of  donor and acceptor as well as their intrinsic radiative lifetime (Sec.~\ref{sec:discussion}).
%%%%%%%%%%%%%%%%%%%%%%%%%%%%%%%%%%%%%%%%%%

%%%%%%%%%%%%%%%%%%%%%%%%%%%%%%%%%%%%%
\section{Calculation of the transfer rates}\label{sec:model}
We consider energy transfer between two emitters in an unbounded dielectric matrix with the permittivity $\epsilon$, located at the points  $\bm r_{\rm D}$ (donor) and $\bm r_{\rm A}=\bm r_{\rm D}+\bm r$(acceptor). The relevant donor and acceptor states are characterized by the energies $E_{\rm D}=\hbar\omega_{\rm D}$ and $E_{\rm A}=\hbar\omega_{\rm A}$ and the transition dipole matrix elements $\bm d_{\rm D}$ and $\bm d_{\rm A}$. In the following we neglect the dispersion and losses in the matrix.
We first present the Fermi Golden rule result for the transfer rate (Sec.~\ref{sec:rule}) and then compare it with the semiclassical Langevin approach (Sec.~\ref{sec:sc1}).
%%%%%%%%%%%%%%%%%%%
\subsection{Fermi Golden rule}\label{sec:rule}
The Fermi Golden rule yields the following expression for the transfer rate
\begin{equation}
\Gamma_{\rm ET,0}=\frac{2\pi}{\hbar}\delta(E_{\rm D}-E_{\rm A})|\bm d_{\rm D}\hat G_{0}\bm d_{\rm A}|^{2}\:, \label{eq:1}
\end{equation}
where
\begin{equation}
G_{0,\alpha\beta}=\frac{3r_{\alpha}r_{\beta}-r^{2}\delta_{\alpha\beta}}{\eps r^{5}}\label{eq:G0}
\end{equation}
is the  electromagnetic Green function evaluated in the electrostatic approximation, and
describing the dipole-dipole coupling between donor and acceptor.\cite{AgranovichGalanin}
This result can be applied for quantum dots as well as molecules. For quantum
dots, we have neglected the local field corrections for the electric
field\cite{lagendijk_review,Poddubny2014arXiv} assuming the permittivities of the dot and
the matrix to be the same. In the case of spherical dots these corrections lead to
renormalization of the dipole matrix element, $\bm d_{\rm D,A}\to \bm d_{\rm
D,A}3\varepsilon_{\rm QD}/(\varepsilon_{\rm QD}+2\eps)$, where $\varepsilon_{\rm QD}$ is
the dot permittivity. In the general case one has to introduce the depolarization factors
depending on the dot orientation and shape.  Additional local field corrections appear for dense arrays of quantum dots.~\cite{Poddubny2014arXiv,Pukhov2008}

The result Eq.~\eqref{eq:1} scales with the distance as $1/r^{6}$. However, Eq.~\eqref{eq:1} neglects any effects of  retardation for the electromagnetic interaction.  When the  retardation effects are taken into account, the transfer rate can be still presented in the form Eq.~\eqref{eq:1}, but the electrostatic potential Eq.~\eqref{eq:G0} should be replaced by the full retarded electromagnetic Green tensor \cite{Dung2002}
\begin{equation}
G_{\alpha\beta}=\left(\delta_{\alpha\beta}+\frac1{q^{2}}\frac{\partial^{2}}{\partial x_{\alpha}\partial x_{\beta}}\right)\frac{e^{\i qr }}{\eps r}\label{eq:G}
\end{equation} evaluated at the transition frequency $\omega=E_{\rm D}/\hbar\equiv E_{\rm A}/\hbar$,
where $q=\omega_{\rm D}\sqrt{\varepsilon}/c$,
so that
\begin{equation}
\Gamma_{\rm ET}=\frac{2\pi}{\hbar}\delta(E_{\rm D}-E_{\rm A})|\bm d_{\rm D}\hat G\bm d_{\rm A}|^{2}\:. \label{eq:2}
\end{equation}
 In this case the long-range radiative transfer, that scales with the distance as $1/r^{2}$, becomes possible.\cite{AgranovichGalanin}
The explicit values for the matrix elements of the interaction $g=\bm d_{\rm D}\hat G_{0}\bm d_{\rm A}$ in the cases, when the dipole momenta of donor and acceptor are parallel to each other and either parallel or perpendicular to the vector $\bm r=\bm r_{\rm D}-\bm r_{\rm A}$ read
\begin{align}
g_{\parallel}=d_{\rm A}d_{\rm D}\frac{\e^{\i qr}}{\varepsilon}\left(\frac{2}{r^{3}}-\frac{2\i q}{r^{2}}\right)\:,\label{eq:gLT}\\
g_{\perp}=d_{\rm A}d_{\rm D}\frac{\e^{\i qr}}{\varepsilon}\left(-\frac{1}{r^{3}}+\frac{\i q}{r^{2}}+\frac{q^{2}}{r}\right)\:.\nonumber
\end{align}
For random mutual orientation of the donor and acceptor matrix elements the transfer is described by the value
$|g^2|=(1/3)|g_{\parallel}^2|+(2/3)|g_{\perp}^2|$, averaged over the orientations.

%%%%%%%%%%%%%%%%%%%%%%%%%%%%%%%%
\subsection{Semiclassical approach}\label{sec:sc1}
Here,  we are going to re-derive Eq.~\eqref{eq:1} within the Langevin random source technique and  the semiclassical theory of light-matter interaction.\cite{Ivchenko2005,Welsch2006,Voronov2007}
%%%%%%%%%%%%%%%
\subsubsection{Radiative decay of the donor}
We start with the radiative decay of the donor state in the absence of acceptors.
The donor  electric polarizability tensor reads
\begin{equation}
\alpha_{\mu\nu} (E)=\frac{d_{D,\mu}d_{D,\nu}}{E_{\rm D}-E},
\end{equation}
i.e. its dipole moment $\bm p_{\rm D}$ induced by the external electric field $\bm E$ at the frequency $\hbar\omega$ is given by
\begin{equation}
\bm p_{\rm D}=\frac{\bm d_{\rm D}\bigl[\bm d_{\rm D}\cdot \bm E(\bm r_{\rm D})\bigr]}{E_{\rm D}-\hbar\omega}\:.\label{eq:D}
\end{equation}
On the another hand, the electric field of the donor is determined by the Green function Eq.~\eqref{eq:G}\:,
\begin{equation}
\bm E(\bm r)=\hat G(\bm r-\bm r_{\rm D})\bm p_{\rm D}\:.\label{eq:E}
\end{equation}
Combining Eq.~\eqref{eq:D} and Eq.~\eqref{eq:E} we obtain the self-consistency condition for the mode of donor coupled with its own electromagnetic field:
\begin{equation}
(E_{\rm D}-\hbar\omega)\bm p_{\rm D}=\bm d_{\rm D} [\bm d_{\rm D}\cdot\hat G(0,\omega)\bm p_{\rm D}]\:.\label{eq:D2}
\end{equation}
This equation allows one to determine the modification of the lifetime of the donor state due to the interaction with light. The energy shift of the donor state can be obtained as well, but this requires regularization of the Green function  taking the finite extent of the wave function into account, see Refs.~\onlinecite{Ivchenko2005,lagendijk_review}. Below we assume, that such regularization has been already performed and is included in the definition of the energy $E_{\rm D}$. We also use the weak coupling approximation when the Green function in the right-hand side of Eq.~\eqref{eq:D2} is evaluated at the frequency $\omega_{\rm D}$.
The spontaneous emission rate is then determined from Eq.~\eqref{eq:D2} as
\begin{equation}
\Gamma_{\rm rad,0}\equiv -2\Im\omega=2d_{\rm D,\alpha}\Im\hat G_{\alpha\beta}(0,\omega_{\rm D}) d_{\rm D,\beta},
\end{equation}
or, explicitly,\cite{Novotny2006}
\begin{equation}
\Gamma_{\rm rad,0}=\frac{4d_{D}^{2}}{3\hbar}\left(\frac{\omega_{\rm D}}{c}\right)^{3}\sqrt{\eps}\:.\label{eq:D0}
\end{equation}
%%%%%%%%%%%%%%%
\subsubsection{Donor decay in the presence of acceptor}

Equation~\eqref{eq:D0} is a textbook result for the spontaneous emission rate.\cite{Novotny2006} However, the approach above can be straightforwardly generalized to include the energy transfer processes.\cite{Shahbazyan2013,Poddubny2015}  To this end,
 Eqs. \eqref{eq:E}--\eqref{eq:D2} should be modified to account for the electromagnetic coupling of the donor and the acceptor as follows:
\begin{align}
&(\omega_{\rm D}-\omega)\bm p_{\rm D}=\frac1{\hbar}\bm d_{\rm D}[\bm d_{\rm D}\cdot(\hat G(0,\omega)\bm p_{\rm D})+
\hat G(\bm r,\omega)\bm p_{\rm A}]\:,\label{eq:syst}\\\nonumber
&(\omega_{\rm A}-\i\gamma_{\rm A}-\omega)\bm p_{\rm A}=
\frac1{\hbar}\bm d_{\rm A}[\bm d_{\rm A}\cdot\hat G(\bm r,\omega)\bm p_{\rm D}]\:.
\end{align}
Here, we include the phenomenological (nonradiative) decay rate $\gamma_{\rm A}$ for the acceptor excited state. The decay is due to the energy relaxation to the lower acceptor states. We are interested in the weak coupling regime, and consider the energy relaxation of the acceptor state to be much faster  than the energy transfer and the radiative decay of both donor and acceptor states.
Hence, the donor lifetime in the presence of acceptor is given by  the perturbative solution of the system Eqs.~\eqref{eq:syst} at the frequency  close to $\omega_{\rm D}$.
The result can be presented as
\begin{equation}
\frac{1}{\tau_{{\rm D},0}}=\Gamma_{{\rm rad},0}\, , \quad \frac{1}{\tau_{\rm D}}=\Gamma_{{\rm rad},0}+\Gamma_{\rm D},\label{eq:tD}
\end{equation}
\begin{equation}
\Gamma_{\rm D}=\frac{2}{\hbar}\Im\left[\frac1{\omega_{\rm A}-\i\gamma_{\rm A}-\omega_{\rm D}}\frac1{\hbar^{2}}\left(\bm d_{\rm D}\cdot\hat G(\bm r)\bm d_{\rm A}\right)^{2}\right]\:.\label{eq:final}
\end{equation}
The second term in Eq.~\eqref{eq:tD} describes the acceptor-induced contribution to the decay rate of the donor state.
This expression is quite different from the standard result Eq.~\eqref{eq:2}. First, Eq.~\eqref{eq:final} includes the finite lifetime of the acceptor state. Second, the functional dependence of Eq.~\eqref{eq:2} and Eq.~\eqref{eq:final} on the (complex) Green function is different. The difference between Eq.~\eqref{eq:2} and Eq.~\eqref{eq:final} constitutes the central result of this work.  Qualitatively, it is due to the fact that Eq.~\eqref{eq:2} describes only the rate of the generation of particles in the acceptor state. On the other hand, Eq.~\eqref{eq:final} is the total acceptor-induced modification of the donor decay rate, which is contributed by both energy transfer to the acceptor and modification of the spontaneous decay rate. In the following Sec.~\ref{sec:discussion} we will analyze
Eq.~\eqref{eq:2} and Eq.~\eqref{eq:final}  in more detail. Here, we only mention that in the case when the distance between the donor and the acceptor becomes much smaller than the wavelength, $qr\ll 1$ and the retardation effects are neglected, Eq.~\eqref{eq:final} reduces to
\begin{equation}
\Gamma_{D,0}= \frac{2\pi}{\hbar}\frac{1}{\pi \hbar}\frac{\gamma_{\rm A}}{(\omega_{\rm D}-\omega_{\rm A})^{2}+\gamma_{\rm A}^{2}}|\bm d_{\rm D}\hat G_{0}\bm d_{\rm A}|^{2} \label{eq:final_{0}}.
\end{equation}
This expression is equivalent to the Fermi Golden rule result Eq.~\eqref{eq:1} in the limit of the vanishing acceptor decay rate. Here we consider only  the case of transparent medium, $\Im \eps=0$. The more general case of lossy medium, where the additional decay channel due to medium heating is  possible, has been analyzed in Ref.~\onlinecite{barnett1996}, see also Ref.~\onlinecite{Glazov2011}.

\subsubsection{Population of acceptors}
In the previous paragraph we have calculated the  decay rates of donor state. Now we will obtain the acceptor population using the same semiclassical technique. To this end, we consider the regime of stationary incoherent pumping and use the random source approach.\cite{Voronov2007}
Hence, the  system Eq.~\eqref{eq:syst} is modified as
\begin{align}
(\omega_{\rm D}-\omega)\bm p_{\rm D}=\frac1{\hbar}\bm d_{\rm D}[\bm d_{\rm D}\cdot(\hat G(0,\omega)\bm p_{\rm D})&+
\hat G(\bm r,\omega)\bm p_{\rm A}]\nonumber\\\nonumber&\qquad+\bm d_{\rm D}\xi(\omega)\:,\nonumber\\\label{eq:syst2}
(\omega_{\rm A}-\i\gamma_{\rm A}-\omega)\bm p_{\rm A}=
\frac1{\hbar}\bm d_{\rm A}[\bm d_{\rm A}\cdot\hat G(\bm r,\omega)&\bm p_{\rm D}]\:,
\end{align}
where $\xi(\omega)$ is the random source term describing the stationary incoherent generation of excitons in the donor state. Generally, the correlations of the random sources are determined by the pumping mechanism,\cite{JETP2009} the simplest approximation
 corresponds to white Gaussian noise
\begin{equation}
\langle \xi^{*}(\omega)\xi(\omega')\rangle=\frac{S}{2\pi}\delta(\omega-\omega'),\quad
\langle \xi^{*}(t)\xi(t')\rangle=S\delta(t-t')
\end{equation}
where $S$ is the exciton generation rate.
First, we calculate the stationary donor state population as
\begin{equation}
N_{\rm D}=\frac{\langle|\bm p_{\rm D}(t)|^{2}\rangle}{|\bm d_{\rm D}|^{2}},
\end{equation}
where
\begin{equation}
\bm p_{\rm D}(t)=\int \frac{\rmd\omega}{2\pi}\bm p_{\rm D}(\omega)\e^{-\i\omega t}\:.
\end{equation}
and
the angular brackets denote the averaging over the random source realizations.
Explicitly, we obtain
\begin{equation}
N_{\rm D}=\left\langle \left|\int \frac{\rmd\omega}{2\pi}\mathcal D_{\rm D}(\omega)\xi(\omega)\e^{-\i\omega t}\right|^{2}\right\rangle,
\end{equation}
where
\begin{equation}
\mathcal D_{\rm D}(\omega)=\frac{1}{\omega_{\rm D}-\omega-\i/(2\tau_{\rm D})}
\end{equation}
is the donor Green function calculated including both energy transfer and radiative decay processes.
The averaging and integration yields
$
N_{\rm D}=S\tau_{\rm D}\:,
$
i.e. the donor population is equal to the product of the lifetime and the generation rate.
The acceptor population is obtained in a similar way as
\begin{equation}
N_{\rm A}\equiv \frac{\langle |\bm p_{\rm A}|^{2}\rangle}{|\bm d_{\rm A}|^{2}}=S|\bm d_{\rm D}\cdot\hat G_{0}\bm d_{\rm A}|^{2} \int \frac{\rmd\omega}{2\pi}
|\mathcal D_{\rm A}(\omega)|^{2}|\mathcal D_{\rm D}(\omega)|^{2}
\end{equation}
with
\begin{equation}
\mathcal D_{\rm A}(\omega)=\frac{1}{\omega_{\rm A}-\omega-\i\gamma_{\rm A}}\:.
\end{equation}
The result of integration reads
\begin{equation}
N_{\rm A}=\frac{2\pi}{\hbar}S\tau_{\rm D}\tau_{\rm A}\frac1{\pi\hbar}\frac{\gamma_{\rm A}}{(\omega_{\rm A}-\omega_{\rm D})^{2}+\gamma_{\rm A}^{2}}|\bm d_{\rm D}\hat G_{0}\bm d_{\rm A}|^{2}\:,\label{eq:NA}
\end{equation}
where $\tau_{\rm A}=1/(2\gamma_{\rm A})$.
It is instructive to rewrite this result  in the form of the kinetic equation for balance of the (nonradiative) decay in the acceptor state and the energy transfer from the donors
\begin{equation}
\frac{N_{\rm A}}{\tau_{\rm A}}=\Gamma_{\rm ET}N_{\rm D}\:.
\end{equation}
Using this equation as the definition of the energy transfer rate $\Gamma_{\rm ET}$, we
find from Eq.~\eqref{eq:NA}
\begin{equation}
\Gamma_{\rm ET}=\frac{2\pi}{\hbar}\Theta|\bm d_{\rm D}\cdot\hat G\bm d_{\rm A}|^{2},\quad
\Theta=\frac1{\pi\hbar}\frac{\gamma_{\rm A}}{(\omega_{\rm A}-\omega_{\rm D})^{2}+\gamma_{\rm A}^{2}}
\:\label{eq:3}
\end{equation}
which is the generalization of Eq.~\eqref{eq:2} to the case of finite acceptor state
lifetime.
Eq.~(\ref{eq:3})
directly corresponds to the expression  commonly used for
the realistic multilevel systems,\cite{Crooker2002,Rodina2015} [e.g. Eq.~(1)
from Ref. \onlinecite{Rodina2015}] with $\Theta$ being the  overlap integral between the donor emission and
the acceptor absorption spectra for the considered model with two-level donor and acceptor.
 We note, that these results can be equivalently obtained using the Keldysh
diagram technique,\cite{Keldysh1965} its correspondence to the Langevin source technique for this problem
is discussed in Refs.~\onlinecite{Voronov2007,JETP2009}.

%%%%%%%%%%%%%%%%%%%%%%%%%%%%%%%%%%%%%%%%%%
\subsection{Ohmic losses approach}
The acceptor excitation rate Eq.~\eqref{eq:3} can be also calculated in a slightly different but equivalent way as  the rate of the absorption of donor emission.\cite{Shahbazyan2011,Agranovich2011} This allows one to interpret the energy transfer process in the form of the Ohmic losses for the donor emission. Thus, one can  separate the contributions to the total donor decay rate Eq.~\eqref{eq:final},  determined by the energy transfer process and by the modification of the far-field emission by the acceptor.

In particular, the acceptor dipole moment induced by the donor with the dipole moment $\bm p_{D}=\bm d_{D}$ is obtained from the second of Eqs.~\eqref{eq:syst} as
\begin{equation}\label{eq:pA}
p_{A,\alpha}=\frac{d_{A,\alpha}d_{A,\beta}G_{\beta\gamma}d_{D,\gamma}}{\hbar (\omega_{\rm A}-\omega_{\rm D}-\i\gamma_{\rm A})}\:,
\end{equation}
and the electric field at the acceptor position is given by
\begin{equation}
E_{\alpha}(\bm r_{\rm A})=G_{\alpha\beta'}(\bm r)d_{D,\beta'}\:.\label{eq:Ea}
\end{equation}
The rate of power absorption is then determined by the standard electrodynamic expression\cite{landau08}
\begin{equation}
\frac{1}{\tau_{\rm ET}}\equiv \Gamma_{\rm ET}=2\Im p_{A,\alpha}E^{*}_{D,\alpha}.\label{eq:4}
\end{equation}
Substituting Eq.~\eqref{eq:pA} and Eq.~\eqref{eq:Ea} into Eq.~\eqref{eq:4} we recover  Eq.~\eqref{eq:3}.

In order to distinguish between the energy transfer and the far-field emission processes we will use the identity\cite{Welsch2006}
\begin{equation}
\int {\rm d}^3 r'' G_{\mu\nu}(\bm r,\bm r'')G^*_{\mu'\nu}(\bm r',\bm r'')\varepsilon''(\bm r'')=4\pi \Im G_{\mu\mu'}(\bm r,\bm r')\:.\label{eq:Gidentity}
\end{equation}
valid for the Green function in arbitrary medium in the case of zero external stationary magnetic field.
For $\bm r=\bm r'$ the right-hand side determines the local density of photonic states and the radiative decay rate.\cite{sprik1996}
For $\mu=\nu',\bm r=\bm r'$ Eq.~\eqref{eq:Gidentity} simplifies to
\begin{equation}
\Im \int {\rm d}^3 r'' \frac{\varepsilon(\bm r'')-1}{4\pi} |G^*_{\mu'\nu}(\bm r'',\bm r)|^{2}=\Im G_{\mu\mu}(\bm r,\bm r)\:.
\end{equation}
The radiative decay rate is due to the far-field emission and due to the Joule heating of the medium.
The Joule losses are determined as the integral in the left hand side over the finite volume, where $\Im \epsilon \ne 0$. The far-field emission is given from the contribution to the integral at $r''\to \infty$ for $\Im \epsilon(\bm r'')\to 0$. For given $\mu$ the integral can be rewritten as $\Im \int d^{3}r''\bm \Pi(\bm r'')\bm E^{*}(\bm r'')$, where
$\Pi$ is the polarizability tensor,
\begin{align}
&\Pi_{\beta}(\bm r'')=\frac{\varepsilon(\bm r'')-1}{4\pi} G_{\beta\alpha}(\bm r'',\bm r),\quad
E_{\beta}(\bm r'')= G_{\beta\alpha}(\bm r'',\bm r)\:.
\end{align}
This expression is equivalent to Eq.~\eqref{eq:4}  and corresponds to the transfer rate
 $\Gamma_{\rm ET}$ in Eq.~\eqref{eq:3}.
The total acceptor-induced  decay rate of the donor state $\Gamma_{\rm D}$ in Eq.~\eqref{eq:final} includes the contribution Eq.~\eqref{eq:3}  due to the Ohmic losses  and the correction to the far-field emission. Thus, the far-field contribution is obtained as the difference between $\Gamma_{\rm D}$ and $\Gamma_{\rm ET}$,
\begin{equation}
\Delta\Gamma_{\rm rad}=\Gamma_{\rm D}-\Gamma_{\rm ET}\:.\label{eq:5}
\end{equation}
%%%%%%%%%%%%%%%%%%%%%%%
\section{Kinetic equations}\label{sec:kinetic}
In the previous section we have presented four approaches yielding consistent results, namely
(i) Fermi Golden rule to calculate  the transfer rate to the acceptor state Eq.~\eqref{eq:1} neglecting the losses and retardation, (ii)
coupled dipole  technique to calculate the donor decay rate Eq.~\eqref{eq:final},
(iii) random sources technique and (iv) Joule power losses approach to calculate the transfer rate for the acceptor state Eq.~\eqref{eq:3}.

These results allow us to formulate the following system of phenomenological kinetic equations determining the population of the donor and acceptor states $N_{\rm D}$ and   $N_{\rm A}$, and the population $N_{{\rm A},0}$ of the acceptor ground (emitting) state which lifetime $\tau_{{\rm  A},0}$ is controlled by the  spontaneous emission:
\begin{align}
\frac{\rmd N_{\rm D}}{\rmd t}&=
-\left(\Gamma_{\rm rad,0}+\Delta\Gamma_{\rm rad}\right)N_{\rm D}-\Gamma_{\rm ET}N_{\rm D}+S\nonumber\\&\equiv
-\frac{N_{\rm D}}{\tau_{\rm D}}+S\:,\nonumber
\\\frac{\rmd N_{\rm A}}{\rmd t}&=-\frac{N_{\rm A}}{\tau_{\rm A}}+\Gamma_{\rm ET}N_{\rm D}\:, \nonumber
\\\frac{\rmd N_{{\rm A},0}}{\rmd t}&=-\frac{N_{{\rm A},0}}{\tau_{{\rm A},0}}+\frac{N_{\rm A}}{\tau_{\rm A}}\:. \label{eq:NA2}
\end{align}
Here, $S$ is the exciton generation rate for the donor state, and the total acceptor-induced correction to the donor decay rate
$\Gamma_{\rm D} = \Gamma_{\rm ET}+\Delta\Gamma_{\rm rad}$  consists of two parts, the correction due to acceptor excitation ($\Gamma_{\rm ET}$ )  and the correction corresponding to the far field emission
($\Delta\Gamma_{\rm rad}$).  In the case of small distance between donor and acceptor the value of $g$ is almost real, $\Gamma_{\rm D}=\Gamma_{\rm A}$ and  $\Gamma_{\rm rad}\ll \Gamma_{\rm D}$.
The expression for $\Gamma_{\rm rad}$ can be explicitly written as
\begin{equation}
\Delta\Gamma_{\rm rad}=-\frac{4}{\hbar}\frac{2(\Im g)^{2}\gamma_{\rm A}+2\Delta \Im (g)\Re g}{\Delta^{2}+\gamma_{\rm A}^{2}}\:,\label{eq:DGamma_rad}
\end{equation}
where $g=\bm d_{D}\cdot\hat G\bm d_{A}$ and $\Delta=\omega_{\rm D}-\omega_{\rm A}$.
Hence, $\Delta\Gamma_{\rm rad}$ is not equal to zero only when the retardation effects are taken into account ($\Im g\ne 0$). This means that the term $\Delta\Gamma_{\rm rad}$ corresponds to the radiation of real photons. On the other hand, $\Gamma_{\rm ET}$ is proportional to $(\Re g)^{2}+(\Im g)^{2}$, i.e. it includes contributions of both real and virtual photons.\cite{Andrews2004}

In the general case the values of $\Delta\Gamma_{\rm rad}$ and $\Gamma_{\rm D}$ can be negative.
For vanishing detuning between donor and acceptor ($\Delta =0$) one has $|\Gamma_{\rm D}|<\Gamma_{\rm ET}$, and $\Delta\Gamma_{\rm rad}<0$, i.e. the far field emission is suppressed, see Eq.~\eqref{eq:DGamma_rad}. For large detuning, ($|\Delta|\gg \gamma_{\rm A}$) the value of $\Gamma_{\rm \rm rad}$ can be positive, i.e. the far field emission is enhanced.
It is also possible, that $\Gamma_{\rm D}$ is equal to zero, but $\Gamma_{\rm ET}$ is not zero. This means that the growth of the donor decay rate due to transfer is exactly compensated by the suppression of the far field emission from the donor.

%%%%%%%%%%%%%%%%%%%%%%%%%%%%%%%%%%%%%%%%%%
\begin{figure}[t!]
\includegraphics[width=0.45\textwidth]{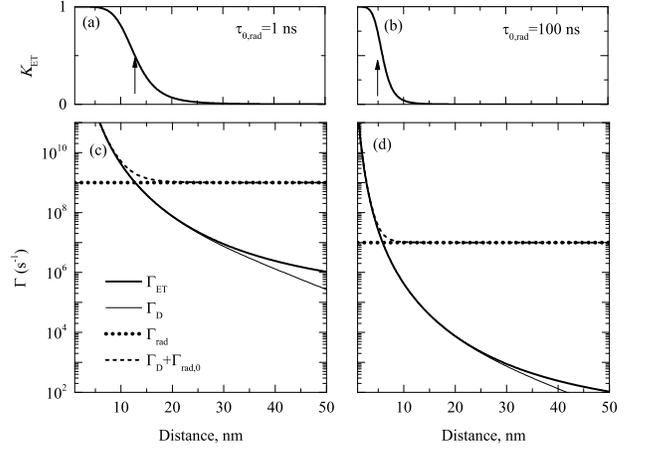}
\caption{Distance dependence of the energy transfer for $\Delta=0$ and $\tau_{\rm rad,0}\approx 1~$ns (a,c) and
$\tau_{\rm rad,0}\approx 100~$ns (b,d).
 Panels (a) and (b) show the transfer efficiency $K_{\rm ET}$, panels (c) and (d) show the full donor decay rates $\Gamma_{\rm D}$ (thin solid curves), energy transfer rates
$\Gamma_{\rm ET}$ (thick solid  curves),  radiative decay rates $\Gamma_{\rm rad}$ (dotted  curves) and the full donor decay rate $\Gamma_{\rm D}+\Gamma_{\rm rad,0}$ (dashed curves).
Other calculation parameters are as follows: $\tau_{A}=1~$ps, $\Delta=0$, $E_{\rm D}=2~$eV, $d_{D}=d_{A}=e\times 0.32$~nm (a,c), $d_{D}=d_{A}=e\times 0.032$~nm (b,d) , $\epsilon_{b}=10$. The rates are averaged over donor and acceptor orientations. Arrows in (a), (b) indicate the F\"orster radii where $K_{\rm ET}=0.5$.
}\label{fig:2}
\end{figure}

We stress that in our model the lifetime of the acceptor excited state $\tau_{\rm A}$ is determined by the nonradiative process and is the shortest time in the system, while the  lifetime of the acceptor ground state $\tau_{{\rm A},0}$ is of the same order as  $\tau_{{\rm D},0}$ so that  $\tau_{\rm A} \ll \tau_{{\rm A},0} \approx  \tau_{{\rm D},0}$. As  discussed above, the donor lifetime $\tau_{{\rm D}}$  can be shortened or lengthened in the presence of acceptor, however the condition $\tau_{\rm A} \ll \tau_{{\rm D}}$ remains valid.

The dynamics of the system Eq.~\eqref{eq:NA} in the absence of stationary pumping for given population of donors at $t=0$  under these conditions is given by
\begin{eqnarray}
N_{\rm D}&=&N_{\rm D}(0)\e^{-t/\tau_{\rm D}},  \label{eq:ND2}\\
 N_{\rm A}&=&\Gamma_{\rm ET}N_{\rm D}(0)\frac{\tau_{\rm A}\tau_{\rm D}}{\tau_{\rm A}-\tau_{\rm D}}(\e^{-t/\tau_{\rm A}}-\e^{-t/\tau_{\rm D}})   \nonumber \\
&\approx&\Gamma_{\rm ET}N_{\rm D}(0)\tau_{\rm A}\e^{-t/\tau_{\rm D}} \, , \nonumber \\
N_{{\rm A},0}&=&\Gamma_{\rm ET}N_{\rm D}(0)\frac{\tau_{{\rm A},0}\tau_{\rm D}}{\tau_{{\rm A},0}-\tau_{\rm D}}\left( \e^{-t/\tau_{{\rm A},0}}-\e^{-t/\tau_{\rm D}} \right) \,\nonumber .
\end{eqnarray}
For stationary pumping the solution of Eqs.~\eqref{eq:NA} reads
\begin{equation}
N_{\rm D}=G\tau_{\rm D},\quad N_{\rm A}=\Gamma_{\rm ET}\tau_{\rm A}N_{\rm D},\quad N_{{\rm A},0} =\Gamma_{\rm ET}\tau_{\rm A,0}N_{\rm D}.
\end{equation}
The acceptor population in the ground (emitting) state can be also rewritten as
\begin{equation}
N_{{\rm A},0}=G\tau_{A,0}K_{\rm ET}\:,
\end{equation}
where
\begin{equation}
K_{\rm ET}=\frac{\Gamma_{\rm ET}}{1/\tau_{\rm D,0}+\Gamma_{\rm D}}=\frac{\Gamma_{\rm ET}\tau_{\rm D,0}}{1+\Gamma_{\rm D}\tau_{\rm D,0}} \:\label{eq:KET}
\end{equation}
is the efficiency of the energy transfer. If we assume that the quantum yield of donor emission without acceptor was equal to one and its intensity was just given by $I_{\rm D}=G$, the modified donor intensity in the presence of acceptor is now given by $I_{\rm D}^*=G(1-K_{\rm ET})$, while the intensity from acceptor is given by $I_{\rm A}^*=G K_{\rm ET}$. It turns out that in the presence of FRET with $\Gamma_{\rm ET}>0$ the quantum efficiency of the donor PL is always decreased even in the case $\Delta \Gamma_{\rm rad}>0$ (increase of the donor radiative rate). However, the increase of the donor radiative rate decreases the efficiency of the energy transfer and vice versa without changing the energy transfer rate $\Gamma_{\rm ET}$.

%%%%%%%%%%%%%%%%%%%%%%%%%%%%%%%%%%%%%%%%%%
\begin{figure}[t!]
\includegraphics[width=0.45\textwidth]{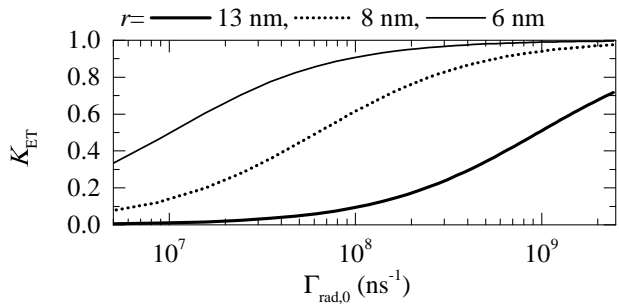}
\caption{
 Dependence of the energy transfer efficiency $K_{\rm ET}$  on the radiative rate  $\Gamma_{\rm rad,0}$ for different donor-acceptor distances $r$ .
Calculation has been performed for $\Delta=0$ and the same other parameters as in Fig.~\ref{fig:2}.
}\label{fig:3}
\end{figure}

%%%%%%%%%%%%%%%%%%%%%%%%%%%%%%%%%%%%%%%%%%
\begin{figure}[t!]
\includegraphics[width=0.45\textwidth]{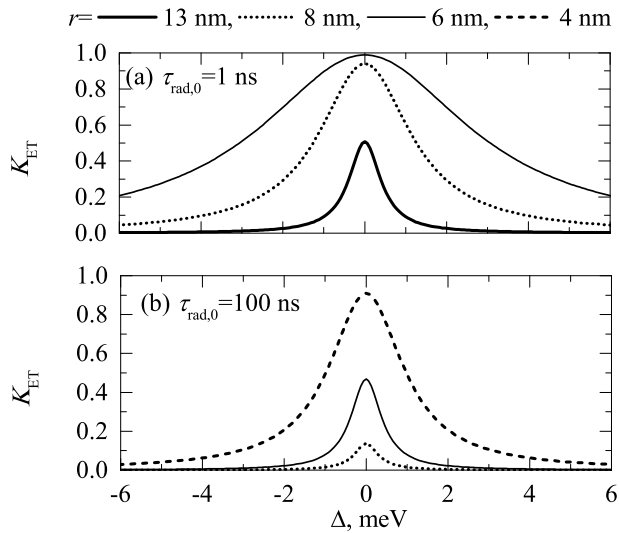}
\caption{
Dependence of the energy transfer efficiency $K_{\rm ET}$ on the donor-acceptor energy detuning $\Delta$  for $\tau_{\rm rad,0}=1~$ns (a) and
$\tau_{\rm rad,0}=100~$ns (b). Thick solid, dotted, thin solid, and dashed curves  correspond to donor-acceptor distance $r=13$~nm, $8~$nm, 6~nm and 4~nm, respectively.
 Other calculation parameters are the same as in Fig.~\ref{fig:2}.}\label{fig:4}
\end{figure}

\section{Results and discussion}\label{sec:discussion}
%%%%%%%%%%%%%%%%%%%%%%%%%%%%%%%%%%%%%%%%%%
Now we proceed to the analysis of the transfer efficiency and transfer rates. We study
their dependence on the donor-acceptor distance $r$ (Fig.~\ref{fig:2}), radiative
lifetimes $\tau_{\rm rad,0}$ (Fig.~\ref{fig:3}) and spectral detunings $\Delta$
(Fig.~\ref{fig:4}).  Figure~\ref{fig:2} examines the distance dependence of the
efficiency $K_{\rm ET}$ (a,b) and the rates $\Gamma_{\rm ET}$, $\Gamma_{\rm D}$, 
$\Gamma_{\rm rad}$,  $\Gamma_{\rm D}+\Gamma_{\rm rad,0}$ (c,d). We have chosen two
respresentative values of the dipole matrix elements $d_{\rm D}=d_{\rm A}$, resulting in
the bare  radiative lifetimes $\tau_{\rm rad,0}=1$~ns
(Fig.~\ref{fig:2}a,Fig.~\ref{fig:2}c) and   $\tau_{\rm rad,0}=100$~ns
(Fig.~\ref{fig:2}b,Fig.~\ref{fig:2}d). The typical  values of the
radiative decay times for the bright exciton in quantum dots may vary from 0.2--0.3~ns to
20~ns depending on the dot type,\cite{Biadala,Patton2003}  while for the  dark quantum dot
exciton transitions the times may vary from 100 ns to 1--2~$\mu$s.\cite{Rodina2015,Patton2003} It has been recently demonstrated that at low temperatures dark excitons determine the energy transfer in dense ensemble of colloidal CdTe nanocrystals.\cite{Rodina2015}
 The nonradiative decay rate of the acceptor state $\tau_{\rm A}$ is equal to $1$~ps.\cite{Klimov2000}
For short radiative lifetime $\tau_{\rm rad,0}=1$~ns the transfer is efficient ($K_{\rm ET}>0.5$) up to the distance $r\approx 13$~nm, which is by definition the radius of the F\"orster process. For larger radius $\Gamma_{\rm ET}$ becomes smaller than $\Gamma_{\rm rad}$ (cf. solid and dotted curves in Fig.~\ref{fig:2}c) and the transfer is suppressed. Comparing thick and thin solid curves in panel (c) one can see, that up to $r\lesssim 40$~nm one has $\Gamma_{\rm ET}\approx \Gamma_{\rm D}$. This means that the transfer is purely nonradiative for $r\lesssim 40$~nm. At larger distances, when the curves deviate, the radiative correction becomes comparable with the transfer rate, although still smaller than $\Gamma_{\rm rad}$. However, at such large distances the transfer is quite inefficient, $K_{\rm ET}\ll 1$. Thus, we conclude from the analysis of Fig.~\ref{fig:2}a and Fig.~\ref{fig:2}c that when the F\"orster process is efficient, it is nonradiative.
For longer radiative lifetime $\tau_{\rm rad,0}=100$~ns (Fig.~\ref{fig:2}b and Fig.~\ref{fig:2}d) the distance dependence of the transfer remains qualitatively the same, but the F\"orster radius shrinks to about $6$~nm.
 The sensitivity of the F\"orster radius to the radiative lifetime reflects the fact that  the radiative rate $\Gamma_{\rm rad,0}$ and the F\"orster rate $\Gamma_{\rm ET}$ are proportional to the second and fourth power of the dipole matrix element, respectively. In Fig.~\ref{fig:2}  the dipole matrix element has been chosen equal for donors and acceptors, $d_D=d_A$, so its increase boosts the relative efficiency of the transfer.
 Hence, in order to enhance F\"orster interaction between the quantum states of the same origin it is beneficial to select the acceptor states with radiative lifetime that is short (but still longer than the nonradiative time $\tau_A$).
The dependence of the F\"orster radius on the radiative lifetime is further analyzed in Fig.~\ref{fig:3}. It shows the transfer efficiencies at different donor acceptor distances as functions of the radiative rate. The calculation confirms that  the transfer at the distances beyond  $10$~nm requires  the radiative lifetime of the acceptor excited state to be as short as 1~ns.
%Figure~\ref{fig:3} (b) shows the distance dependence of the rates at a given radius $r=8$~nm. Clearly, for long radiative  lifetime ($\Gamma_{\rm rad,0}<1/100$~ns$^-1$) the radiative contribution to the donor decay rate starts being important and being affected by the acceptor. However, at the same time the transfer process becomes infavorable.

Finally, in Fig.~\ref{fig:4} we present the dependence of the transfer efficiency $K_{\rm ET}$ on the energy detuning between donor and acceptor $\Delta$ for different distances $r=13$~nm, $8~$nm, $6$~nm, and $4$~nm (thick solid, dotted, thin solid, and dashed curves, respectively) and for two different radiative lifetimes $\tau_{\rm rad,0}=1$~ns (a) and
$\tau_{\rm rad,0}=100$~ns (c). The transfer efficiency is a Lorentzian function of the detuning  with maximum at $\Delta=0$. For $\tau_{\rm rad,0}=1$~ns (a) the spectral range of the transfer is on the order of meV and increases at smaller donor-acceptor distances. For long radiative lifetime $\tau_{\rm rad,0}=100$~ns the spectral range strongly decreases and the transfer becomes possible only for the detuning less than $1~$meV and donor-acceptor distance $r\lesssim 5$~nm. The detuning range allowing for  the transfer is also inversely proportional to the nonradiative lifetime of the acceptor state $\tau_A=1/2\gamma_A$, directly entering the overlap integral $\Theta$ in Eq.~\eqref{eq:3}.

%\begin{figure}
%\centering\includegraphics[width=0.45\textwidth]{gD0}
%\caption{Distance dependence of the full donor decay rate (a), FRET rate (b) and radiative part of the decay (c) for different mutual polarizations of the donor and acceptor dipole momenta. }\label{fig:all}
%\end{figure}

%%%%%%%%%%%%%%%%%%%%%%%%%%%%%%%%%%%%%%%%%%
\section{Summary}
To summarize, we have presented a theory of the F\"orster interaction, accounting both for the transfer of the energy from the donor to the acceptor (F\"orster effect) and for the antenna-like modification of the far-field donor emission by the acceptor (Purcell effect). We have demonstrated for  typical parameters corresponding to the semiconductor quantum dots that the Purcell effect is negligible provided that the transfer efficiency is high, $K_{\rm ET}>0.5$. In another words, the fast transfer is purely nonradiative. The radiative corrections start to play role only at relatively large distances $r>40$~nm when the transfer is quenched. We have analyzed the dependence of the F\"orster  radius on the radiative lifetime and revealed that the radius above 10~nm can be achieved only utilizing bright donor and acceptor excitonic states with the radiative lifetime on the order of $1~$ns. In this case the transfer takes place provided that the detuning between donor and acceptor  does not exceed several meV.

While our theory is quite general,  it should be stressed that the numerical results above are applicable only to the transfer in the homogeneous  dielectric matrix. The competition between radiative and nonradiative transfer mechanisms in the case of structured electromagnetic environment (plasmonic\cite{Tumkur2015,Zubin2015} or dielectric\cite{Wubs2015}) requires further studies.

\section*{Acknowledgements}
Stimulating discussions with V.Y. Aleshkin and M.A. Noginov are gratefully acknowledged.
This work has been funded by  the  Russian Science Foundation grant No.14-22-00107.
A.N.P. acknowledges the support of the ``Dynasty'' Foundation.
%\bibliography{FRET}

\end{document}